\documentclass[preprint,12pt,sort&compress]{elsarticle}

\usepackage{amssymb}
\usepackage{amsthm}

\usepackage{lineno}

\usepackage{graphicx}
\graphicspath{ {./Images/} }
\usepackage{amsmath}
\usepackage[version=4]{mhchem}
\usepackage{siunitx}
\usepackage{longtable,tabularx}
\usepackage{xcolor}
\usepackage{float} 
\usepackage{caption}
\usepackage{subcaption}
\usepackage{hyperref}
\usepackage{upgreek}
\usepackage{booktabs}
\usepackage{framed} 
\usepackage{multicol} 
\usepackage{multirow}
\usepackage{nomencl} 
\makenomenclature
\usepackage{etoolbox}
\renewcommand\nomgroup[1]{%
 \item[\bfseries
 \ifstrequal{#1}{L}{Latin Symbols}{%
 \ifstrequal{#1}{M}{Greek Symbols}{%
 \ifstrequal{#1}{N}{Abbreviations}{}}}%
]}

\journal{Acta Astronautica}

\begin{document}

\begin{frontmatter}



\title{Wind--Pellet Shear Sailing}


\author{Jeffrey K. Greason \fnref{JKG}}

\address{Electric Sky Inc., Midland, Texas, 79701, USA}
\address{Tau Zero Foundation, Broomfield, Colorado, 80038, USA}

\author{Dmytro Yakymenko \fnref{DY}}
\author{Mathias N. Larrouturou \fnref{ML}}
\author{Andrew J. Higgins \fnref{AJH}}

\address{Department of Mechanical Engineering, McGill University, Montreal, Quebec, H3A 0C3, Canada}

 
 

\fntext[JKG]{Board Chairman (Tau Zero Foundation), Chief Technologist (Electric Sky, Inc.). Email address: \href{mailto:jeff@greason.com}{\underline{jeff@greason.com}}}
\fntext[DY]{Undergraduate Student}
\fntext[ML]{Undergraduate Student}
\fntext[AJH]{Professor. Email address: \href{mailto:andrew.higgins@mcgill.ca}{\underline{andrew.higgins@mcgill.ca}}}
\begin{abstract}
A propulsion concept in which a spacecraft interacts with both a stream of high-velocity macroscopic pellets and the mass of the interplanetary or interstellar medium is proposed. Unlike previous pellet-stream propulsion concepts, the pellets are slower than the spacecraft and are accelerated backwards as they are overtaken by it, imparting a forward acceleration on the spacecraft. This maneuver is possible due to the interaction with a fixed medium (e.g., interstellar medium); as the spacecraft travels through the medium, it is able to extract power from the relative wind blowing over the spacecraft. As viewed from the rest frame, the kinetic energy of the pellets is transferred to the spacecraft; the source of energy for the propulsive maneuver (i.e., whether the source is the pellets or the wind) is thus reference-frame dependent. This concept relies upon the relative velocities (or shear) between the pellet stream and the fixed medium in order to concentrate the energy of the pellets into the spacecraft and is therefore termed wind--pellet shear sailing. The equations governing the mass ratio of pellets to the spacecraft and its dependence on the final spacecraft velocity are derived, analogous to the classical rocket equation; the critical role of the efficiency of the power extraction and transfer process is identified. Natural sources of energy (photon and particle fluxes from the sun) are considered as a means to accelerate the pellets, produced from \emph{in-situ} sources of material in the solar system, to velocities of 1000 to 6000~km/s. Techniques for onboard generation of power via electromagnetic interaction with the interstellar medium (ISM) are reviewed, with a repetitively stroked plasma magnet being identified as a promising approach. The necessity of the spacecraft to detect and track the pellets as they are overtaken dictates the desired properties of the pellets. Pellet pushers (i.e., accelerators) on board the spacecraft are also preliminarily explored in their engineering considerations, with electric field accelerators of charged nanometric particles, Lorentz-force accelerated ionized pellets, or expansion of vaporized pellets via a nozzle being highlighted as potential approaches. A preliminary mission profile is defined in which a 500-kg scientific payload is delivered to orbit about $\upalpha$-Centauri, using wind--pellet shear sailing as the intermediate stage to bring the spacecraft from 2\% to 5.5\% of $c$, with a total mission duration (launch to arrival) of 27~years. The concept design illustrates the potential for synergy at the low-velocity end with advanced solar photon or solar-wind sailing and at the high-velocity end with the \emph{q}-drive concept.

\end{abstract}




\end{frontmatter}


\nomenclature[L#01]{$c$}{speed of light [m/s]}
\nomenclature[L#02]{$D$}{drag on spacecraft due to interaction with interstellar medium [N]}
\nomenclature[L#03]{$h$}{flow enthalpy [J/kg]}
\nomenclature[L#04]{$L_\mathrm{acc}$}{length of accelerator [m]}
\nomenclature[L#05]{$m_\mathrm{sc}$}{mass of spacecraft [kg]}
\nomenclature[L#06]{$m_\mathrm{p}$}{mass of individual pellet [kg]}
\nomenclature[L#07]{$\dot{m}_\mathrm{ISM}$}{mass flux of interstellar medium that the spacecraft interacts with [kg/s]}
\nomenclature[L#08]{$\dot{m}_\mathrm{p}$}{mass flux of pellets encountered by spacecraft [kg/s]}
\nomenclature[L#09]{$m_{\mathrm{p}_\mathrm{tot}}$}{total mass of pellets [kg]}
\nomenclature[L#10]{$P_{\mathrm{ext}}$}{Power extracted from interstellar medium [W]}
\nomenclature[L#12]{$q$}{electric charge on particle [C]}
\nomenclature[L#13]{$T$}{thrust on spacecraft due to pellet acceleration [N]}
\nomenclature[L#14]{$T_\mathrm{net}$}{net thrust on spacecraft [N]}
\nomenclature[L#15]{$U_\mathrm{ISM}$}{velocity of interstellar medium in spacecraft frame [m/s]}
\nomenclature[L#16]{$U_\mathrm{p}$}{velocity of pellets in spacecraft frame [m/s]}
\nomenclature[L#17]{$U_\mathrm{1}$}{velocity of pellets entering nozzle [m/s]}
\nomenclature[L#18]{$U_\mathrm{2}$}{velocity of pellets exiting nozzle [m/s]}
\nomenclature[L#19]{$V$}{voltage [V]}
\nomenclature[L#20]{$v_\mathrm{sc}$}{velocity of spacecraft in rest frame [m/s]}
\nomenclature[L#21]{$v_\mathrm{p}$}{velocity of pellets in rest frame [m/s]}
\nomenclature[L#22]{$v_\mathrm{sc}$}{velocity of spacecraft in rest frame [m/s]}
\nomenclature[M#01]{$\alpha$}{ratio of pellet mass to ISM mass}
\nomenclature[M#02]{$\eta$}{net efficiency}
\nomenclature[M#03]{$\eta_\mathrm{ext}$}{efficiency of power extraction from ISM}
\nomenclature[M#04]{$\eta_\mathrm{acc}$}{efficiency of power transmission to pellets}
\nomenclature[M#08]{$\xi$}{non-dimensional distance of spacecraft travel}
\nomenclature[M#09]{$\phi$}{non-dimensional velocity of spacecraft}

\printnomenclature[1.5cm] 


\section{Introduction}
\label{sec:intro}
\noindent
Reaching velocities useful for practical interstellar flight is fundamentally a problem of supplying sufficient kinetic energy to the spacecraft. A 1000-kg interstellar probe (comparable to the mass of the Voyager spacecraft) at 20\% of the speed of light carries $1.8 \times 10^{18}$~J of kinetic energy. However it is acquired, this value is the equivalent of converting 20~kg of rest~mass to energy, utterly beyond the reach of chemical energy, and requiring implausible mass ratios even for advanced fusion rockets (e.g., mass ratio of 160~000 for an exhaust velocity of $5 \times 10^6$~m/s). Of course, the much greater masses involved in speculative future human missions to the stars make matters even more difficult.

Recent work has suggested a spacecraft, interacting with the surrounding interstellar medium (ISM), may be able to extract power via this interaction in order to accelerate onboard reaction mass \cite{Greason2019}. This mechanism provides a means of concentrating the kinetic energy of a larger reaction mass carried on board into the smaller mass of the spacecraft itself via a process resembling inelastic collisions of the reaction mass with the surrounding medium. This approach provides some relief from the ``tyranny of the rocket equation,'' in that a second stage employing the technique combined with an advanced fusion rocket as the first stage could, in principle, reduce the overall mass ratio to $\approx 500$, but this is still a challenging requirement.

Beaming momentum to a vehicle via photons \cite{Marx1966, Forward1984, Lubin2016} or particles \cite{Landis2004, Landis2001} reflected from a photon or magnetic sail has long been recognized as a means to circumvent the limitations imposed by carrying reaction mass. The challenge of beamed momentum is twofold. First, the range of the beam must be very long to reach practical interstellar velocities, implying an enormous aperture for a photon beam or a small dispersion for a particle beam. Second, the power supplied to the beam remains large (even larger than that transferred to the spacecraft, due to various inefficiencies in the creation and transmission of the beam), and the cost of the beam transmitting facility tends to scale strongly with beam power.

The problem of dispersion can be addressed by the use of macroscopic pellets \cite{Singer1980} potentially enhanced by onboard course correction to further reduce dispersion \cite{Nordley1993, Kare2001, Kare2002}. The problem of the energy and expense of the pellet launching system remains daunting, with the length of accelerator required to launch macroscopic pellets to 20\%~of~$c$ being on the order of $10^4$~km \cite{Peaslee1979, Higgins2018}. Kare later proposed laying down a ``runway'' of fusion-fuel pellets that would be moving slower than the spacecraft \cite{Greason2020a}, which reduces the mass ratio compared to a pure rocket system and greatly lessens the demands placed on the accelerator.

Clearly, in facing the central problem of collecting and concentrating the energy necessary for interstellar propulsion, it is attractive to convert the energy available from natural sources in space directly into a form useful for spacecraft propulsion. This is the motivation behind solar-photon sails \cite{McInnes1999, Davoyan2021} and solar-wind particle sails \cite{Zubrin1991, Zubrin2000, Janhunen2004, Slough2007}. In these cases, the ``beam source'' is provided for free by the Sun, but in both cases, there are speed limits ($< 0.02 \, c$) well below those necessary for practical interstellar flight.

If we use these natural energy sources (the solar photon flux or solar wind) to launch macroscopic pellets, the pellets are traveling well below the desired final speed of the spacecraft. Macroscopic pellets can provide a long range as their dispersion due to thermal effects is negligible, but in this scenario, the spacecraft is overtaking the pellets and cannot be pushed by them.\footnote[1]{We are aware of a prior proposal from Lebon \cite{Lebon1986} that involves a spacecraft being accelerated to velocities necessary for interstellar flight as it traverses over a stream of pellets moving at lower speed. The concept involves a spacecraft with an onboard solenoid that overtakes a stream of paramagnetic pellets; since magnetic fields are conservative, the mechanism by which this technique generates forward thrust on the spacecraft is unclear to the authors.} Those pellets, however, are still carrying a flux of kinetic energy when viewed from the rest frame (i.e., a frame of reference at rest with respect to the interstellar medium (ISM)). If we use the pellet stream to transmit \emph{energy} to the spacecraft rather than \emph{momentum}, they can supply that energy even though they are moving slower than the spacecraft. This technique exploits the velocity difference, or \emph{shear}, between the pellets and the ISM, in a manner analogous to the way ocean sailing vessels exploit the velocity difference between the wind and the ocean to gain energy, and thus is termed \emph{wind--pellet shear sailing}. In traditional sailing, the difference between the direction of the wind and the direction of travel of the sailboat requires the use of a keel to interact with a second medium---the water---so that the resultant force is in the direction of travel. With wind--pellet shear sailing, the interactions are unidirectional, exploiting the relative motion of the pellets and the wind, which are both moving in the same direction as viewed from the spacecraft-fixed reference frame. This approach offers the possibility of using a stream of inexpensive pellets, accelerated by harvesting natural energy sources, and then concentrating that kinetic energy into a spacecraft for propulsive purposes, without the need to carry the reaction mass on board the spacecraft.

Wind--pellet shear sailing is closely analogous to a technique used terrestrially that has generated a considerable controversy in recent decades: sailing straight downwind faster than the wind. The principle that enables a vehicle to be propelled in the downwind direction while it exceeds the wind speed relies on the vehicle exploiting the difference in relative velocities between the wind and the vehicle and between the wind and the fixed ground (or water). A recent embodiment of this concept is the \emph{Blackbird}, an engineless cart with a propeller connected to the rear axle by a transmission chain \cite{Muller2021a, Muller2021b}. The cart starts from rest, being pushed by the wind and accelerated up to the speed of the wind. Then, it can accelerate further because the rotation of the wheels by the ground inputs power into the cart, which is transmitted to the propeller. The propeller pushes the air backwards and creates forward thrust. As viewed by a fixed observer, the wind, after going through the propeller, is slowed down relative to the ground. From an energy perspective, the kinetic energy that was initially in the wind has thus been transferred to the cart, accelerating it further. The full optimization of a ground vehicle exploiting the difference in speed between the wind and ground was explored in detail by \cite{Bauer1969}; for a recent review of the history of this problem, see \cite{McDonald2021}. In relation to the concept explored in the present paper, the pellets play the role of the wind while the ISM plays the role of the fixed ground: the spacecraft overtakes the pellets and---as viewed from the rest frame---decelerates them to zero, transferring their energy to the spacecraft. The power input necessary to effect this pellet deceleration is extracted from an interaction with the ISM.

\section{Analysis}
\label{sec:Analysis}

As the concept of wind--pellet shear propulsion is new, appreciation of the technique benefits from a hierarchy of analyses wherein increasingly realistic factors are taken into consideration. In the following subsections, progressively detailed analyses of the concept are presented.

\begin{figure}
\centering
\includegraphics[width=0.75\textwidth]{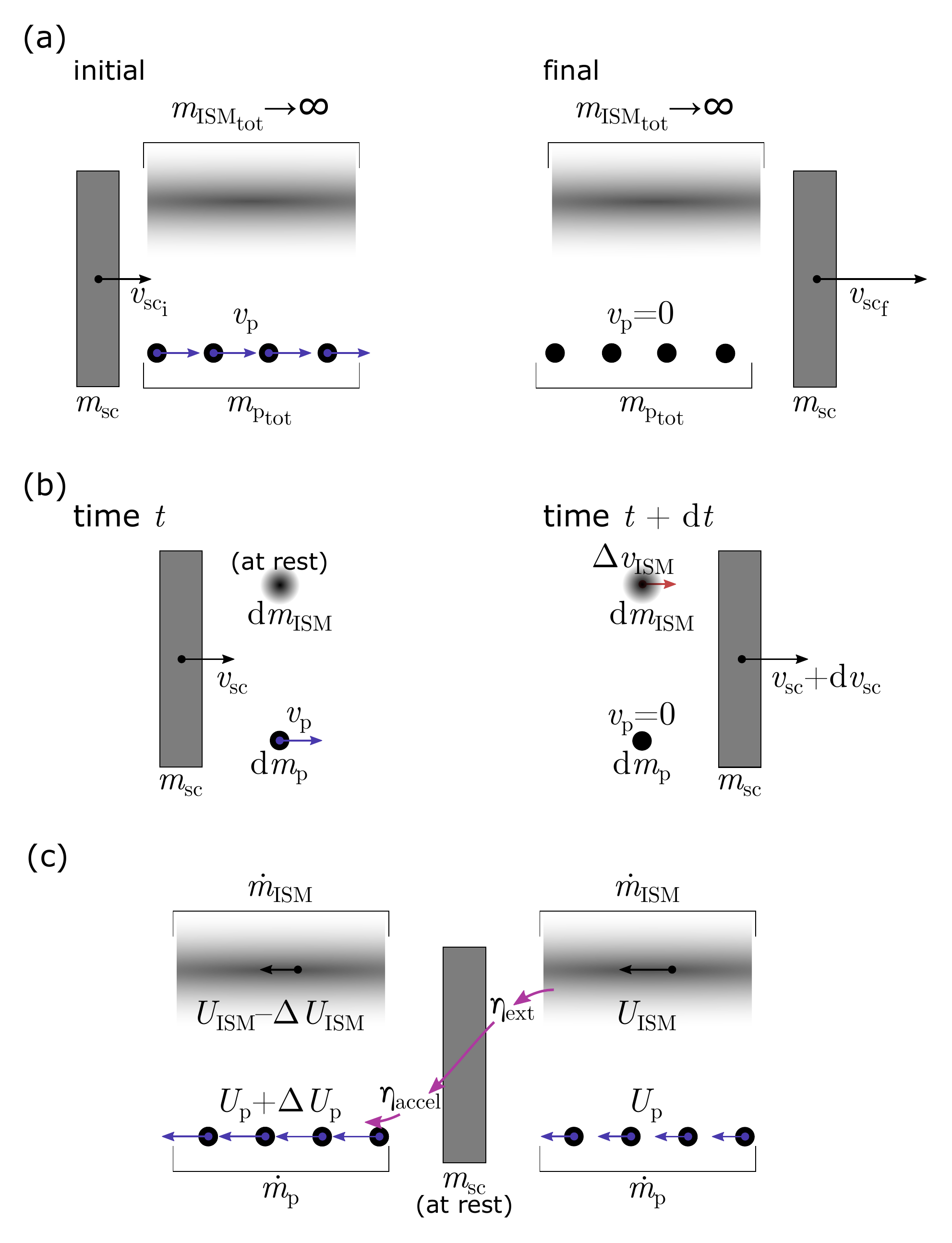}
\caption{Schematic of wind--pellet shear sailing, defining different terms used in analysis. \textbf{(a)} For the case of infinite ISM mass and perfect efficiency, all of the energy of the pellets is transferred to the spacecraft (viewed from inertial frame). \textbf{(b)} Differential interaction with a finite mass of ISM, with motion $\Delta V_\mathrm{ISM}$ induced in the ISM (viewed from inertial frame). \textbf{(c)} View from spacecraft-fixed reference frame, showing efficiency $\eta_\mathrm{ext}$ of power extraction and efficiency $\eta_\mathrm{acc}$ of pellet acceleration.}
\label{fig:Schematic}
\end{figure}

\subsection{Infinite ISM limit}\label{subsec:InfiniteISM}
In an implementation that is easily understood, wind--pellet shear propulsion would involve a spacecraft overtaking a stream of pellets moving at $v_\mathrm{p}$ as viewed in the rest frame. As the spacecraft overtakes the pellets, it accelerates them backwards, bringing the pellets to zero velocity in the rest frame. This condition (i.e., leaving the pellets at rest following the spacecraft) can be shown to give the optimal transfer of energy to the spacecraft: If the pellets retained residual velocity, their motion would represent energy that could have been used by the spacecraft but was wasted.

If all interactions between the spacecraft and pellets (see schematic in Fig.~\ref{fig:Schematic}(a)) are conservative, the conservation of energy is
\begin{equation}\label{eq:EnergyInifiteISM}
\frac{1}{2} \, m_\mathrm{sc} v_\mathrm{sc_\mathrm{i}}^2 + \frac{1}{2} \, m_\mathrm{p_{tot}} \, v_\mathrm{p}^2 = \frac{1}{2} \, m_\mathrm{sc} v_\mathrm{sc_\mathrm{f}}^2 
\end{equation}
and solving for the final spacecraft velocity yields
\begin{equation}\label{eq:FinalVelocity1InifiteISM}
v_\mathrm{sc_\mathrm{f}} = \sqrt{\frac{m_\mathrm{p_{tot}}}{m_\mathrm{sc}} v_\mathrm{p}^2 + v_\mathrm{sc_\mathrm{i}}^2 }.
\end{equation}
Momentum is conserved by interacting with the infinite mass  of the ISM. In the limit of an interaction with a large mass of ISM, most of the pellet momentum has been transferred to the ISM, but in the limit of infinite ISM, the velocity induced in the ISM approaches zero.

In this subsection and subsequent analyses presented below, it is assumed that the spacecraft has already achieved the pellet velocity as it initiates the wind--pellet shear propulsive maneuver ($v_\mathrm{sc_i} = v_\mathrm{p}$); it is likely that the same technique used to accelerate the pellet dispenser would also be used to accelerate the spacecraft to the same starting velocity. In this case, the final spacecraft velocity simplifies to
\begin{equation}\label{eq:FinalVelocity2InifiteISM}
v_\mathrm{sc_\mathrm{f}} = v_\mathrm{p} \sqrt{\frac{m_\mathrm{p_{tot}}}{m_\mathrm{sc}} + 1}.
\end{equation}
%
%
The resulting Eq.~(\ref{eq:FinalVelocity2InifiteISM}) is the analog to the rocket equation. This relation conveys the core of the wind--pellet shear propulsion concept and demonstrates a principal advantage over conventional rocket propulsion: the required reaction mass (i.e., the pellets) increases \emph{quadratically} with the desired final velocity of the spacecraft, rather than \emph{exponentially} as it does with a conventional rocket. This relation will also be seen to comprise a bounding limit for the more detailed solutions with finite ISM and efficiencies included.

\subsection{Finite ISM limit viewed from rest frame}\label{subsec:FiniteISM}

The next level of analysis considers the case where the mass of ISM that the power extraction device (i.e., the windmill on board the spacecraft) interacts with is finite, rather than infinite as considered above. The momentum loss of the spacecraft (i.e., drag) is no longer neglected in this analysis. Examining a differential process in which the spacecraft interacts with a differential element of the pellet stream $\mathrm{d}m_\mathrm{p}$ and a differential element of the surrounding medium $\mathrm{d}m_\mathrm{ISM}$ (see Fig.~\ref{fig:Schematic}(b)), the conservation of momentum
\begin{equation}
m_\mathrm{sc} \, v_\mathrm{sc} + \left (\mathrm{d} m_\mathrm{p} \right) v_\mathrm{p} = m_\mathrm{sc} \, \left ( v_\mathrm{sc} + \mathrm{d} v_\mathrm{sc} \right) + \left ( \mathrm{d}m_\mathrm{ISM} \right) \Delta v_\mathrm{ISM} 
\end{equation}
can be solved for the velocity change induced in the ISM ($\Delta v_\mathrm{ISM}$, in the same direction as the spacecraft)
\begin{equation}\label{eq:DeltaVmomentum}
\Delta v_\mathrm{ISM} = \left ( v_\mathrm{p} - m_\mathrm{sc} \frac{\mathrm{d} v_\mathrm{sc}}{\mathrm{d} m_\mathrm{p}}\right ) \left (\frac{\mathrm{d} m_\mathrm{p}}{\mathrm{d} m_\mathrm{ISM}} \right ).
\end{equation}
The conservation of energy after the interaction is
\begin{equation}
\frac{1}{2} m_\mathrm{sc} \, v_\mathrm{sc}^2 + \frac{1}{2} \left (\mathrm{d}m_\mathrm{p} \right) \, v_\mathrm{p}^2 = \frac{1}{2} m_\mathrm{sc} \, \left ( v_\mathrm{sc} + \mathrm{d} v_\mathrm{sc} \right )^2 + \frac{1}{2} \left ( \mathrm{d}m_\mathrm{ISM} \right ) \left (\Delta v_\mathrm{ISM} \right )^2.
\end{equation}
Neglecting terms of $\left (\mathrm{d}v_\mathrm{sc} \right )^2$ that appear in isolation and using (\ref{eq:DeltaVmomentum}), the following equation of motion can be found
\begin{equation}\label{eq:FiniteISMode}
\alpha \, m_\mathrm{sc}^2 \left ( \frac{ \mathrm{d} v_\mathrm{sc} }{ \mathrm{d} m_\mathrm{p} } \right )^2 + 2 \, m_\mathrm{sc} \, \left ( v_\mathrm{sc} - \alpha v_\mathrm{p} \right ) \left ( \frac{ \mathrm{d} v_\mathrm{sc} }{ \mathrm{d} m_\mathrm{p} } \right ) + \left ( \alpha - 1 \right ) v_\mathrm{p}^2 = 0
\end{equation}
where
\begin{equation}
\alpha = \frac{m_\mathrm{p}}{m_\mathrm{ISM}}
\end{equation}
is the ratio of the pellet mass processed by the spacecraft to the ISM mass encountered by the spacecraft. This equation cannot be solved analytically, but it can be nondimensionalized and solved numerically (see Appendix~A) to build up a series of solution curves for constant values of $\alpha$, giving the final spacecraft velocity in terms of the pellet velocity and the pellet-to-spacecraft mass ratio, see Fig.~\ref{fig:VelocityCurves}. As shown in Appendix~B, wind--pellet shear sailing is capable of generating positive acceleration of the spacecraft only for the case where $\alpha < 1$, i.e., the local mass of the ISM that the spacecraft interacts with must exceed the mass of pellets that are processed over the same interval.

\begin{figure}
\centering
\includegraphics[width=0.95\textwidth]{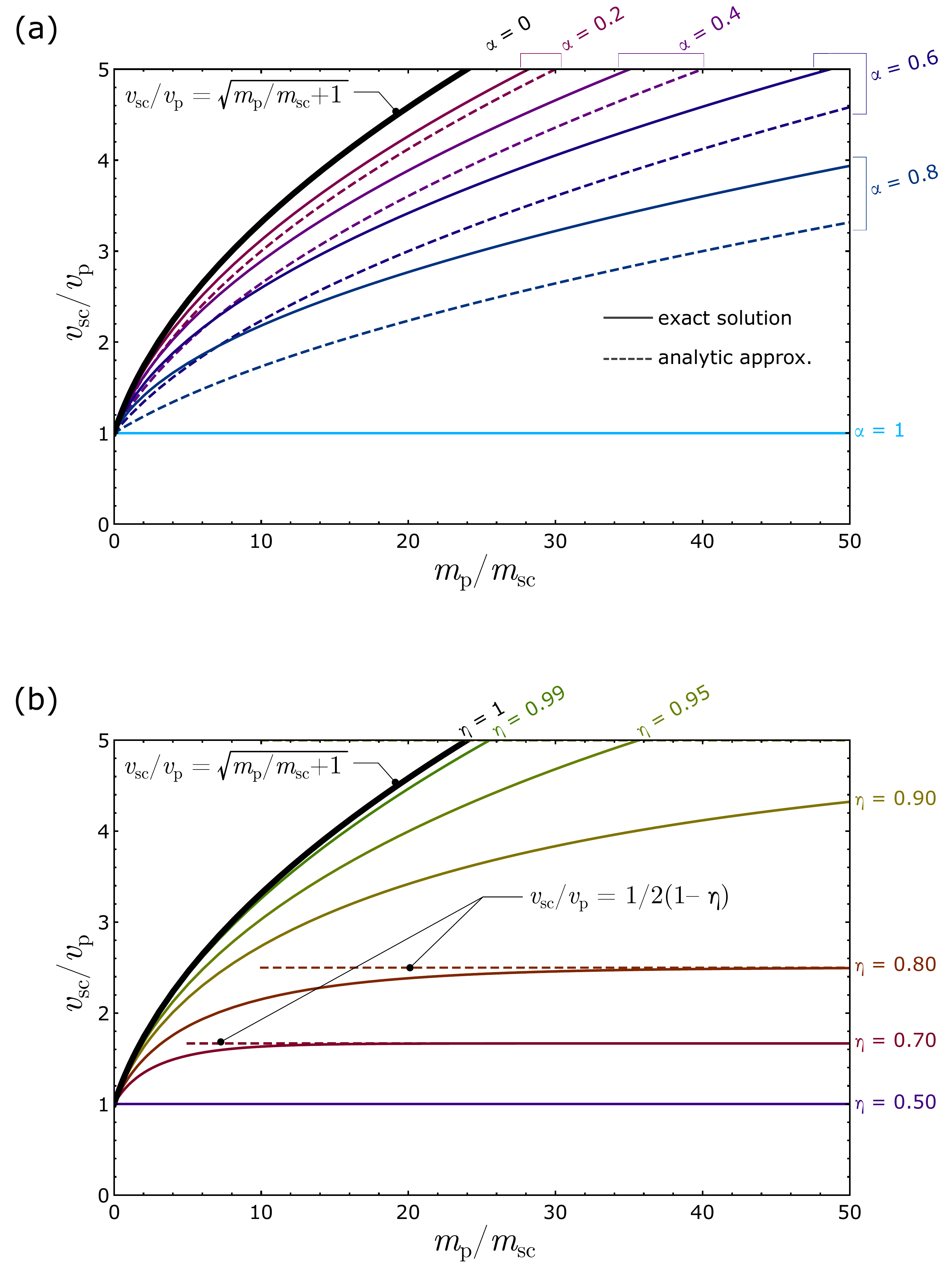}
\caption{Spacecraft velocity (nondimensionalized by pellet velocity) as a function of pellet-to-spacecraft mass ratio. \textbf{(a)} Effect of interacting with finite pellet-to-ISM mass ratio $\alpha$ (efficiency $\eta = 1$ for this plot). Dashed curves are from the analytic approximation Eq.~\ref{eq:FinalVelocityFiniteISM}. \textbf{(b)} Effect of power extraction and acceleration efficiency $\eta$ (pellet-to-ISM mass ratio $\alpha = 0$ for this plot). Dashed lines are the limit to velocity as the pellet-to-spacecraft mass ratio approaches infinity Eq.~\ref{eq:VelocityLimit}).}
\label{fig:VelocityCurves}
\end{figure}

Note that in the limit of $\alpha \rightarrow 0$ (i.e., the mass of ISM that the spacecraft interacts with is infinite), Eq.~(\ref{eq:FiniteISMode}) reduces to
\begin{equation}\label{eq:FiniteISModeAlphaZero}
m_\mathrm{sc} \, v_\mathrm{sc} \,\mathrm{d} v_\mathrm{sc} = \frac{1}{2} \, v_\mathrm{p}^2 \, \mathrm{d} m_\mathrm{p}
\end{equation}
which can be solved via separation of variables to produce the same result as Eq.~(\ref{eq:FinalVelocity2InifiteISM}), demonstrating that the results of this subsection, when taken to the limit of infinite ISM mass, correspond to those in Subsection~\ref{subsec:InfiniteISM}.

While the more general equation (\ref{eq:FiniteISMode}) cannot be solved analytically, a closed-form solution can be found in the limit of small $\alpha$ via a Taylor series expansion, as shown in Appendix~B, the result being

\begin{equation}\label{eq:FinalVelocityFiniteISM}
v_\mathrm{sc_\mathrm{f}} \approx v_\mathrm{p} \sqrt{\left ( 1 - \alpha \right ) \frac{m_\mathrm{p_{tot}}}{m_\mathrm{sc}} + 1}.
\end{equation}
In comparing this result to Eq.~(\ref{eq:FinalVelocity2InifiteISM}), the effect of a finite ratio to pellet mass to ISM mass ($\alpha$) is seen to have the effect of reducing the energy transfer from the pellets by a factor of $\left ( 1 - \alpha \right)$. This approximate solution is compared to the exact numerical solution in Fig.~\ref{fig:VelocityCurves}(a). The approximate solution given by Eq.~(\ref{eq:FinalVelocityFiniteISM}) is seen to under-predict the final spacecraft velocity and thus forms a conservative bound on the performance of wind--pellet shear propulsion for the case of interaction with a finite mass of ISM.

\subsection{Spacecraft frame with power efficiency}\label{sec:SpacecraftFrame}

The analysis of the increase in spacecraft velocity can also be performed in the spacecraft-fixed frame. While this analysis is performed in a noninertial frame, the mass of the spacecraft remains constant during acceleration, and thus, once the net thrust on the spacecraft is found, the acceleration and final spacecraft velocity can be found via integration of kinematic relations. Note in the analysis that follows, $U$ is used to denote velocities observed from the spacecraft-fixed reference frame, while $v$ is retained to denote velocities observed from the rest (inertial) frame. The spacecraft-fixed frame also is convenient for including terms that include the efficiency of extraction of power from the wind of ISM blowing past the spacecraft $ \eta_\mathrm{ext}$ as follows:
\begin{equation}\label{eq:ExtractionEfficiency}
P_\mathrm{ext} = \eta_\mathrm{ext} \left ( \frac{1}{2} \, \dot{m}_\mathrm{ISM} \, U_\mathrm{ISM}^2 - \frac{1}{2} \, \dot{m}_\mathrm{ISM} \, \left ( U_\mathrm{ISM} - \Delta U_\mathrm{ISM} \right ) ^2 \right ).
\end{equation}
This extracted power is then used to accelerate the pellets with an efficiency $\eta_\mathrm{acc}$ in conversion to kinetic energy of the pellets as follows:
\begin{equation}\label{eq:AccelerationEfficiency}
\eta_\mathrm{acc} P_\mathrm{ext} = \frac{1}{2} \, \dot{m}_\mathrm{p} \, \left ( U_\mathrm{p} + \Delta U_\mathrm{p} \right ) ^2 - \frac{1}{2} \, \dot{m}_\mathrm{p} \, U_\mathrm{p}^2 .
\end{equation}
Note the efficiencies are defined such that $\eta_\mathrm{ext}< 1$ and $\eta_\mathrm{acc}< 1$. When substituting Eq.~(\ref{eq:ExtractionEfficiency}) into Eq.~(\ref{eq:AccelerationEfficiency}), the combined efficiency is denoted $\eta =\eta_\mathrm{ext} \, \eta_\mathrm{acc}$. Following a similar development as in Subsection~\ref{subsec:FiniteISM}, $\Delta U_\mathrm{ISM}$ can be solved for.

The thrust on the spacecraft resulting from the backward acceleration of the pellets is given by
\begin{equation}
T = \dot{m}_\mathrm{p} \, \Delta U_\mathrm{p}
\end{equation}
and the drag associated with the deceleration of the ISM is given by
\begin{equation}
D = \dot{m}_\mathrm{ISM} \, \Delta U_\mathrm{ISM}.
\end{equation}
The net thrust (thrust minus drag), which determines spacecraft acceleration, is thus
\begin{equation}\label{eq:NetThrustFma}
T_\mathrm{net} = m_\mathrm{sc} \frac{\mathrm{d} v_\mathrm{sc}}{\mathrm{d} t} = T - D = \dot{m}_\mathrm{p} \, \Delta U_\mathrm{p} - \dot{m}_\mathrm{ISM} \, \Delta U_\mathrm{ISM}.
\end{equation}
Using the expression for $\Delta U_\mathrm{ISM}$ from energy conservation in (\ref{eq:NetThrustFma}) and transforming back to the rest frame via the following transformation
\begin{eqnarray}
\Delta U_\mathrm{p} = v_\mathrm{p} \\
U_\mathrm{ISM} = v_\mathrm{sc} \\
U_\mathrm{p} = v_\mathrm{sc} - v_\mathrm{p}
\end{eqnarray}
yields the equation of motion
\begin{equation}\label{eq:SCframeEfficienciesODE}
m_\mathrm{sc} \, \frac{\mathrm{d} v_\mathrm{sc}}{\mathrm{d} t} = \dot{m}_p \, v_\mathrm{p} - \dot{m}_\mathrm{ISM} \, v_\mathrm{sc} \left ( 1 - \sqrt{1 - \frac{\left (2 \left(v_\mathrm{sc} - v_\mathrm{p} \right) v_\mathrm{p} + v_\mathrm{p}^2 \right ) \dot{m}_\mathrm{p}}{\eta \, \dot{m}_\mathrm{ISM}\, v_\mathrm{sc}^2}} \right ).
\end{equation}
This equation can be shown to be identical to Eq.~(\ref{eq:FiniteISMode}) for the case of ideal efficiency ($\eta = 1$). The general form of Eq.~(\ref{eq:SCframeEfficienciesODE}) can be nondimensionalized and solved numerically, as done in Appendix~A.

The effect of introducing power handling efficiency on the final spacecraft velocity is shown in Fig.~\ref{fig:VelocityCurves}(b). Note that less-than-perfect efficiency results in the appearance of a maximum velocity relative to the pellet velocity that can be obtained via wind--pellet shear sailing. For the case of interaction with an effectively infinite mass of ISM ($\alpha \rightarrow 0$), it can be shown that the limiting velocity is
\begin{equation}\label{eq:VelocityLimit}
\lim_{m_\mathrm{p} \to \infty} \left ( v_\mathrm{sc_\mathrm{f}} \right)_\mathrm{\alpha \to 0} = \frac{v_\mathrm{p}}{2(1-\eta)} .
\end{equation}
%
This result for a limit velocity is plotted as dashed lines in Fig.~\ref{fig:VelocityCurves}(b). For example, if the power handling efficiency is 75\% ($\eta = 0.75$), the maximum spacecraft velocity is only twice the pellet velocity. This result emphasizes the need for careful consideration of component efficiencies in the technical implementation of wind--pellet shear sailing.

\section{Implementations}

To implement wind--pellet shear sailing, the system has three essential components:
\begin{enumerate}
 \item A \textbf{pellet dispenser}, which creates the runway of pellets moving through the fixed medium at the pellet speed $v_p$
 \item A \textbf{windmill} on board the spacecraft, which extracts energy from the passage of the spacecraft through the medium
 \item A \textbf{pellet pusher} on board the spacecraft, which takes the energy from the windmill to accelerate the pellets backwards, as viewed from the spacecraft frame of reference
\end{enumerate}
In this Section, we will discuss examples of possible implementations of these subsystems. In no case are these intended to be more than conceptual approaches to illustrate that these three subsystems have physically plausible paths towards a later engineering implementation.

\subsection{Pellets and pellet dispensers}

\begin{figure}[H]
 \centering
\includegraphics[width=1.0 \textwidth]{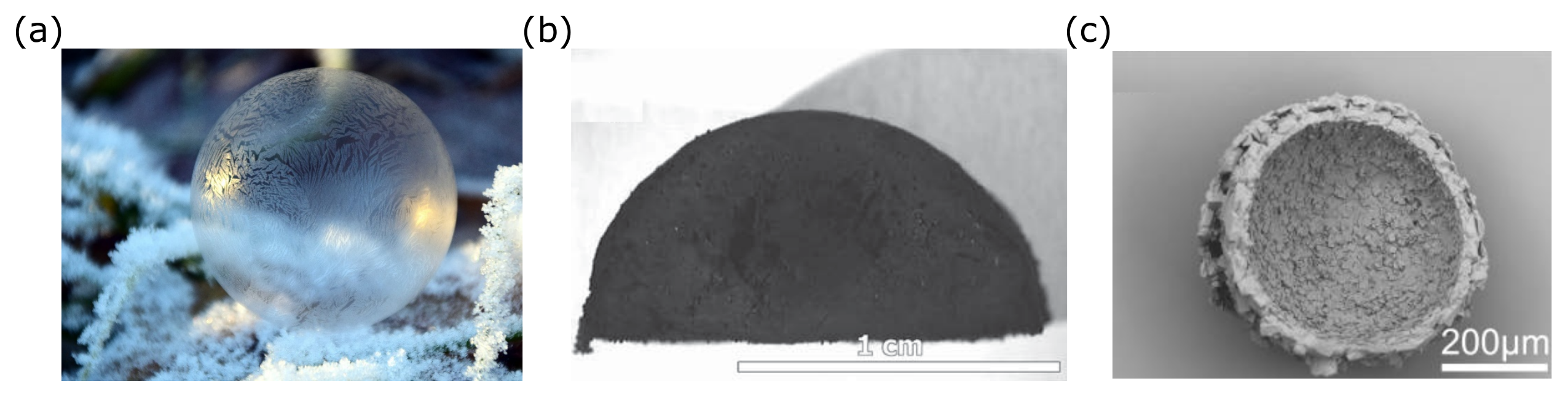}
\caption{Concepts for pellets for use in wind--pellet shear sailing. (a)~Frozen soap bubble, (b)~aerographite \cite{Mecklenburg2012}, (c)~spherical shell agglomerate of azithromycin \cite{Chen2019}.}
\label{fig:Pellets}
\end{figure}

A pellet stream, deployed to be intercepted by a following spacecraft, must have low dispersion, i.e., a minimal random distribution of velocity. For macroscopic pellets, this is a function of the precision of the dispenser rather than a limitation of fundamental physics; this problem was surveyed by Kare \cite{Kare2002}. Drift and accuracy will be the factors influencing dispersion, requiring the following spacecraft to maneuver to keep the intake of the pusher centered on the pellet stream. To do so, the pellets must be large enough for the spacecraft to see them; this works best for low density pellets (e.g., hollow spheres) of larger diameter that are conductive or carry a thin conductive coating so as to maximize their cross section for radar or lidar detection. The expected range of pellet masses that meet these criteria for detectability are in the 0.1--1~mg range, corresponding to hollow pellets of cm scale. Some candidate pellets are illustrated in Fig.~\ref{fig:Pellets}.

Next, we consider how to accelerate the pellets or \emph{pellet dispenser}. One attractive energy source for pellet acceleration is the solar wind. While in the plane of the ecliptic, solar-wind speeds of 400--450~km/s are common; above 20$^\circ$ of solar latitude, wind speeds increase to the 650--750~km/s range \cite{Miralles2011, Verscharen2019}. Many nearby stars lie in this region \cite{RECONS2012}. In this case, a dispensing craft could be accelerated to solar-wind speed by use of a plasma magnet sail \cite{Slough2005, Slough2007} carrying a supply of pellets. The pellet mass in this case might be water (available at many locations in the Solar system \emph{in situ}), formed into frozen bubbles and coated with nanometer-scale metallic films or doped for conductivity. The wall thickness of water bubbles is of order 1~$\upmu$m, providing 0.3~mg at 1~cm diameter. In order to disperse the pellets over the required distance and desired spacing, it might be necessary to release the pellets during the acceleration phase of the dispenser craft, so that early pellets are slower than later pellets, and, over time, they spread out to achieve the runway length required. The same effect can be achieved by putting technically credible fission or fusion propulsion on the dispenser craft to allow some increase in velocity during the dispensing maneuver.

Greater velocities are obviously attractive, and given a means of interacting with the plasmas in space at high lift-to-drag ratio, \emph{dynamic soaring} maneuvers are possible. Recent work by the present authors \cite{Larrouturou2021} suggests speeds up to 2\%~of~$c$ can be achieved in this way, and this technique could be invoked for the dispensing craft. Note that creating the runway via the dispensing craft requires the velocity of the dispenser to change during the process, otherwise the pellets would all stay with the dispenser; given the velocities involved, this may involve using drag to slow down the dispenser while ejecting pellets, favoring multiple smaller dispenser craft launched over time rather than one large one.

An alternative energy source is the photon flux near the sun. Hollow aerographite pellets offer a low effective density and, if released close to the sun, can reach up to 2\%~of~$c$ \cite{Heller2020}. These pellets would again be hollow spheres. To provide the emplacement of the pellets along the direction of the target star, one approach is a \emph{statite} dispensing craft \cite{Forward1991} that hovers near the sun at the dispensing radius using heat shields comparable to the Parker Solar Probe \cite{Fox2016}. This is challenging for either solar sails or plasma magnet sails, but potentially achievable.

\subsection{Windmills}

Due to the low density of the interstellar or interplanetary medium, extracting useful energy by purely mechanical means such as a turbine is impractical. A wind-energy extraction system can be derived from otherwise drag-producing electric or magnetic fields by arranging the field configuration to move downwind repeatedly at a velocity less than the wind, the wind thus doing work on the translating field. See \cite{Greason2019} for a more extensive discussion of this strategy for realizing a practical windmill.

In order to produce useful power extraction, the fields must be physically much larger than the practical dimensions of the spacecraft. This requirement implies that the dominant field configuration interacting with the wind must be a dipole configuration, since field components greater than dipole will become negligible at radii that are large multiples of the spacecraft dimension. This constraint still permits configurations based on oscillating large magnetic fields induced in plasma currents, such as the plasma magnet \cite{Slough2005}, or on large scale electric fields \cite{Janhunen2004}. If the spacecraft velocity vector crosses local magnetic field lines, electrodynamic tethers can also be used for power generation \cite{WentzelLong2020}.

\subsection{Pellet pushers}\label{subsection:Pushers}

If the entire kinetic energy of the pellets is transferred to the spacecraft as they are overtaken by the spacecraft, the pellets should be left at the rest velocity of the inertial (ISM-fixed) frame, as discussed in Subsection~\ref{subsec:InfiniteISM}. Thus, the pellets must undergo a velocity change comparable to their initial velocity, i.e., in the range of 1000--6000 km/s, which sets the requirement for the pellet pusher. Three potential approaches are identified here; we emphasize that these may not be the most promising implementations but are offered as examples.

\begin{figure}
 \centering
\includegraphics[width=0.55\textwidth]{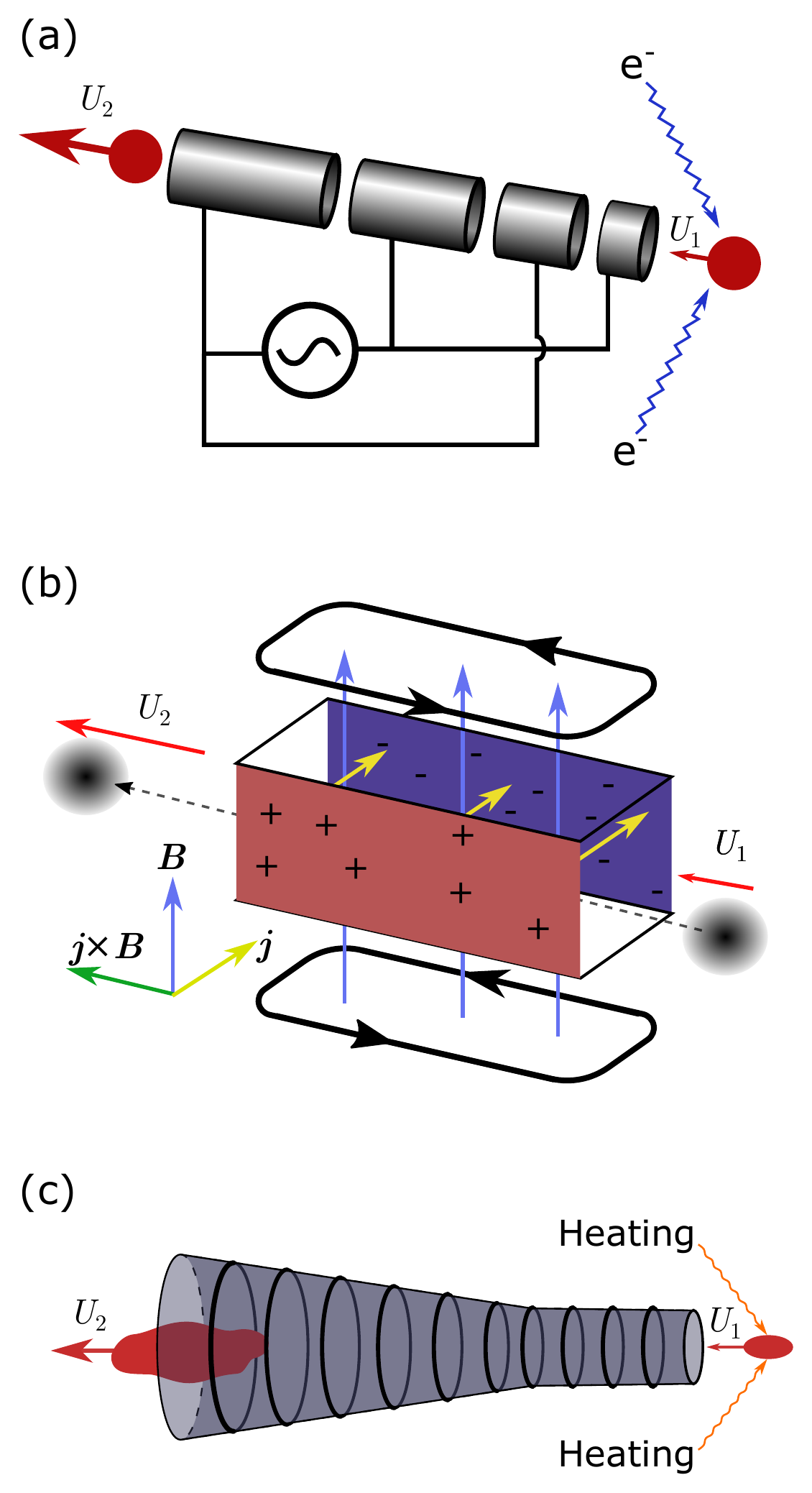}
\caption{Schematics of possible accelerator technologies that could be used to accelerate pellets relative to spacecraft (pellet pushers). (a)~Electric accelerator (linac), (b)~Lorentz accelerator ($J \times B$), (c)~Supersonic nozzle (pellet scramjet).}
\label{fig:Pushers}
\end{figure}

\subsubsection{Electric accelerators}

Dust accelerators employing a single voltage potential (e.g., Van de Graaff, Pelletron, or Cockcroft–-Walton voltage sources) that are capable of accelerating macroscopic pellets on the order of 10~nm in size to velocities of 100~km/s have existed since the early 1960s \cite{Friichtenicht1962}. Multi-stage dust accelerators, which would permit the voltage potential to be re-used, have been proposed \cite{Vedder1978}, but their application to date has been limited. There is no fundamental limitation to the velocities that can be obtained in such an accelerator and operating them in the naturally exquisite vacuum of interstellar space offers a number of advantages, in principle enabling them to be a low-mass component on the spacecraft. A sample implementation---based on an RF-pumped linac---is shown in Fig.~\ref{fig:Pushers}(a). As they approach the spacecraft, pellets would be charged via an onboard electron gun, making the pellets responsive to the applied electric fields of the accelerator.

The governing relation for a multistage electric accelerator is
\begin{equation}\label{eq:ElectriccAccelerator}
\left (\frac{q}{m} \right )_\mathrm{p} \, \left ( \frac{\mathrm{d}V}{\mathrm{d} x} \right)_\mathrm{avg} \, L_\mathrm{acc} = \frac{1}{2} \, \left ( U_2^2 - U_1^2 \right )
\end{equation}
where $\left ( \frac{\mathrm{d}V}{\mathrm{d} x} \right)_\mathrm{avg}$ is the average potential gradient along the accelerator of length $L_\mathrm{acc}$, and $U_1$ and $U_2$ are the initial and final velocity of the pellets relative to the spacecraft, respectively. Expressed in terms of the pellet ($v_\mathrm{p}$) and spacecraft velocities ($v_\mathrm{sc}$) in the rest frame for the case wherein the pellets are brought to rest, Eq.~(\ref{eq:ElectriccAccelerator}) becomes
\begin{equation}
\left (\frac{q}{m} \right )_\mathrm{p} \, \left ( \frac{\mathrm{d}V}{\mathrm{d} x} \right)_\mathrm{avg} \, L_\mathrm{acc} = \frac{1}{2} \, v_\mathrm{p} \left (2 \, v_\mathrm{sc} - v_\mathrm{p} \right ) \approx v_\mathrm{p} \, v_\mathrm{sc}
\end{equation}
where the approximate equality applies in the case where the spacecraft greatly exceeds the pellet velocity, which will be the case for the interstellar flight applications of this approach. Note that parameters (i.e., accelerator length, voltage gradient, charge-to-mass ratio, pellet and spacecraft velocities) are linearly related.

The average potential gradient is likely limited to values of 10~MV/m due to field emission leading to breakdown. If the accelerator length is limited to 100~m, then the required charge-to-mass ratio of the pellets would need to be on the order of $(q/m)_\mathrm{p}$~$\sim$~$10^4$--$10^5$~C/kg. A macroscopic object of milligram mass, previously having been identified as the desirable mass range for pellets, cannot support this charge-to-mass ratio due to limitations imposed by material strength \cite{Higgins2018}. Thus, it is unlikely that pellets can be accelerated while intact; frangible pellets that can be broken down upon electrostatic charging into nanometric particles is a means to overcome this limit.

One possibility for a pellet that can be reduced to smaller particles would be a weakly bound agglomerate. Agglomerates comprised of nanometric powder can be easily and reproducibly made at mm-scale by straightforward means \cite{Mi2013}. This technique could likely be extended to make mm-scale spherical shells of nanometric particles. Carbon fullerenes ($\mathrm{C}_{60}$) would be a promising candidate for the particles that would comprise the agglomerate. Efficient particle acceleration via electrostatic forces is aided by a uniform charge-to-mass ratio of the particles. $\mathrm{C}_{60}$ has the advantage of only existing in a small number of charge states. Doubly charged fullerenes ($\mathrm{C}_{60}^{2+}$) have been successfully accelerated via tandem electrostatic accelerators with 30~MV total potential to velocities of nearly 3000 km/s \cite{Attal1993}---comparable to the velocity change required for wind--pellet shear sailing. Doubly ionizing the $\mathrm{C}_{60}$ molecules contained in a pellet requires on the order of only 1~J of energy, however, the electrostatic energy associated with this surplus charge in a confined space (i.e., the size of the pellet) can be much greater than the initial kinetic energy of the pellet. Thus, it would be necessary to charge the pellet fragments as they expand spatially into the accelerator. If the charged particles are contained and directed aft via magnetic confinement (e.g., a magnetic mirror), this electrostatic energy is converted into directed kinetic energy and is not lost.

\subsubsection{Lorentz force accelerators}

For mg-mass pellets, if the pellet is being tracked and can be targeted by a laser or electron beam, it takes on the order of 100~J of pulse energy to convert the pellet into a plasma that is conductive (15~000~K), which is negligible compared to the kinetic energy contained in a pellet moving at 2\%~of~$c$ as viewed from the rest frame. When the expanding plasma fireball fills a duct on board the vehicle (assumed 1~m$^2$ area), for an ion density $\approx 5 \times 10^{19}$~m$^{-3}$, the duct can apply voltage in one axis perpendicular to the ionized pellet motion and a magnetic field in the other axis, also perpendicular to the pellet motion, inducing Lorentz ($J \times B$) forces to accelerate the plasma. In its simplest form---the crossed-field accelerator---this implementation is a rectangular duct in which current passes between two sides across a perpendicular magnetic field, as shown in Fig.~\ref{fig:Pushers}(b) \cite{Blackman1963}.

%
%

For the case of a mg-mass pellet approaching the duct at 0.03$c$ (in the ship frame of reference) and departing at 0.05$c$, the requirements are challenging, with pulsed currents in the MA range and crossed fields in the range of several T. These currents and fields represent a significant scaling from currently demonstrated systems, but it appears to be a physically plausible approach to a pellet pusher. Note that at these high velocities, magnetohydrodynamic (MHD) approaches become highly efficient, as nearly all the voltage applied appears as Hall voltage, producing an accelerating effect that is on the order of 10$^5$ times greater than the voltage required to overcome resistive losses.

\begin{figure}[H]
 \centering
\includegraphics[width=0.55\textwidth]{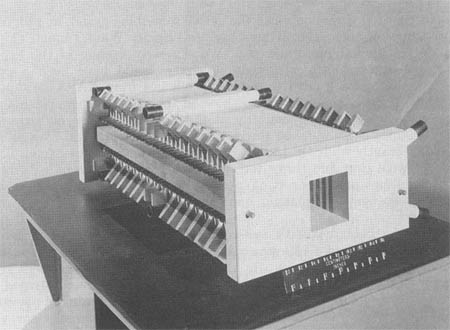}
\caption{20~MW crossed-field plasma accelerator \cite{Hansen1995}.}
\label{fig:LorentzAccelerator}
\end{figure}

\subsubsection{Pellet scramjet}

If the pellet ionization by the on board laser or electron beam is sufficient to elevate the sound speed of the resultant plasma fireball to a value comparable to (but less than) the pellet velocity itself, then it is possible to expand and accelerate the plasma via nozzle expansion. This approach is akin to a scramjet, in which the reaction mass remains supersonic relative to the vehicle while energy is deposited into the flow, with net thrust being obtained upon expansion via a nozzle.

For an adiabatic expansion, the conservation of energy is
\begin{equation}
2 \left (h_1 - h_2 \right ) = 2 \, c_p \left (T_1 - T_2 \right ) = U_2^2 - U_1^2 = \, v_\mathrm{p} \left (2 \, v_\mathrm{sc} - v_\mathrm{p} \right ).
\end{equation}
For a fully ionized pellet (assumed hydrogen) approaching the duct at 0.03$c$ (in the ship frame of reference) and departing at 0.05$c$ (the ship speed in rest frame), the initial temperature entering the nozzle would need to be 0.3~MeV greater than the temperature of the exhaust, giving an inlet Mach number of 1.3. If we set the exhaust Mach number to 20, the exhaust temperature is 3.5 keV (negligible compared to inlet temperature). The nozzle area ratio would be approximately 500. If the pellet were a greater $Z$ material (e.g., water, carbon, etc.), the temperature required increases to several MeV and radiation losses via bremsstrahlung likely become excessive.

A more benign solution may be an \emph{isothermal} expansion, where energy continues to be added to the flow as it expands via laser or RF heating to maintain a constant temperature \cite{Seifert1952}. For an isothermal expansion, the area ratio required is given by
\begin{equation}
\frac{A_2}{A_1} = \exp \left ( \frac{U_2^2 - U_1^2}{2\,\frac{R}{MW}\,T} - \ln{\frac{U_2}{U_1}} \right ).
\end{equation}
In this case, a nozzle area ratio of 100 can be used and the temperature required is lower, approximately 0.15~MeV, with the inlet and outlet Mach numbers being 1.9 and 3.1, respectively. The energy necessary to heat the pellet material to these temperatures is comparable to the kinetic energy contained in the pellet itself (as viewed from the rest frame), so it is essential that most of the thermal energy be converted into directed kinetic energy. The high-temperature exhaust for the isothermal case thus represents a significant energy loss, so likely some combination of isothermal and adiabatic expansion would be used to keep temperatures reasonable and avoid losing thermal energy in the exhaust.

\section{Candidate mission}\label{sec:Mission}

With the physical principles having been established above, we proceed to an illustrative example to show the potential of this technique for interstellar flight. The design is conceptual only, and we arbitrarily set the dry mass of each stage (excluding payload) equal to its payload. Since the power flows are large within the wind--pellet shear sailing stage, how small the dry mass can be made is an important parameter that awaits a more detailed design, although since some options involve all-superconducting machinery for both the windmill and pellet pusher, there is some reason to hope these losses may be made small compared to the power flow.

Consider a spacecraft delivered to $\upalpha$-Centauri of 1000-kg mass after propellant depletion (500-kg scientific payload and 500-kg structure, dry tanks, etc.), comparable to previous flagship interplanetary probes and sufficient to incorporate instruments, a power supply, and a high-bandwidth optical communication system for returning data to Earth. The spacecraft is propelled to its 20\%~of~$c$ cruise velocity by a \emph{q}-drive stage \cite{Greason2019}, which also serves as a plasma-magnet drag device to decelerate the spacecraft upon approaching the target star, permitting an orbiter rather than a flyby mission {\cite{Jackson2020}}. Assuming 95\% efficiency (with 5\% energy loss by heating of the interstellar medium due to parasitic drag of the windmill), a \emph{q}-drive-powered final stage with mass ratio of 15 (15~000-kg wet mass) accelerates the vehicle from 5.5\%~of~$c$ to 20\%~of~$c$.

The earlier stage of propulsion invoking wind--pellet shear sailing, under the same assumptions, has a 15~000-kg mass for the propulsion hardware of that stage and hence a 30~000-kg total mass at the end of the wind--pellet shear sailing maneuver. For simplicity of analysis, consider the pellets to be launched by photon pressure from a dispenser in close proximity to the Sun, for example, as aerographite hollow spheres \cite{Heller2020}. This approach gives a 2\%~of~$c$ pellet speed and using the same assumption of 5\% energy loss to parasitic drag, $\approx$240,000~kg of pellets would be required to accelerate the stage from 2\%~of~$c$ to 5.5\%~of~$c$. If dynamic soaring at the solar-wind termination shock is used \cite{Larrouturou2021}, similar results would be obtained by a series of dispenser craft.

For the wind--pellet shear sailing stage, the spacecraft starts at the same speed as the majority of the pellets; the initial pellets are slightly slower to overcome the starting problem. The spacecraft accelerates along a trajectory with the power passing through the spacecraft limited to a value of 250~GW in order to keep power management workable, accelerating for $\approx$1.4 years to reach 5.5\%~of~$c$, over a runway $\approx$3850~AU long. Placing the pellets over that runway takes $\approx$3.0 years for the first pellets launched to reach that distance, which means the first pellets must be launched $\approx$1.6 years prior to the ship. Assuming the \emph{q}-drive stage takes $\approx$4 years for its acceleration, and the plasma magnet brake takes $\approx$1 year to decelerate, the mission profile is summarized in Table~\ref{tab:MissionProfile}.

\begin{table}
\centering
\caption{Mission Profile}
\label{tab:MissionProfile}
\begin{tabular}{@{}lrl@{}}
\toprule
Event & Time\\
\midrule
\small{ Pellet dispenser launch } & \small { T+0 years} \\
\small{ Spacecraft stack launch (dynamic soaring \cite{Larrouturou2021} to reach 2\%~$c$) } & \small { T+0 years } \\
\small{ Pellet dispenser at Sun begins dispensing } & \small { T+1 year } \\
\small{ Spacecraft begins wind--pellet sailing } & \small { T+2.6 years } \\
\small{ Spacecraft overtakes end of pellet runway } & \small { T+4.0 years } \\
\small{ \emph{q}-drive acceleration finished and begin coast at 20\% $c$ } & \small { T+8.0 years } \\
\small{ Begin plasma-magnet breaking } & \small { T+26.3 years } \\
\small{ Braking complete } & \small { T+27.3 years } \\
\small{ Arrive in $\upalpha$-Cen orbit} & \small { T+27.3 years } \\
\small{ First orbiting data returned } & \small { T+31.6 years } \\
\bottomrule
\end{tabular}
\end{table}

While development cost and the cost of the unique spacecraft hardware cannot be estimated at this early stage, the aggregate mass launched is on the order of 300~000 kg, the majority of which is inert reaction mass to be formed into the pellets by the dispenser craft. At current launch prices to LEO (assuming a low-thrust high-$I_\mathrm{sp}$ electric propulsion stage is used for Earth escape) of \$2000/kg, the launch cost would be on the order of \$600~million---by no means prohibitive for a flagship mission of this type. In the future, with larger missions, either lower cost launch or materials derived from \emph{in-situ} sources in space could be used, which might reduce prices as low as \$100/kg for bulk mass---opening up the possibility of even greater mass interstellar missions, including perhaps, in the further future, human missions.

By way of comparison, direct scaling of the laser-propelled photon sail approach of the type currently contemplated for g-mass-scale spacecraft for a 500-kg payload would, at current \$100/W prices, have an acquisition cost of \$$10^{17}$, and even at the hoped-for future price of \$1/W for laser photons, an acquisition cost of \$$10^{15}$ (i.e., a quadrillion dollars). While these are acquisition costs and not per-mission costs, the price ratio between these two approaches of order $10^7$ illustrates why the pellet-sailing approach, harvesting natural energies, is so attractive in comparison to laser propulsion. In essence, it converts the sun into a long-range pellet propulsion system harvesting gigawatts of source power, over years, to be concentrated in the payload’s kinetic energy upon passage of the mission spacecraft.

\section{Discussion}

In this paper, we have put considerable emphasis on the pellet dispenser and pellet pusher because we were initially uncertain that they could be physically realized. However, the portion of the implementation needing the most development is the windmill. The windmill must collect large power flows from a diffuse wind and do so with high efficiency and low parasitic drag. While the pusher concepts discussed in Subsection~\ref{subsection:Pushers} have been previously published, more detailed and practical implementations are a fruitful area for further research.

The design of a statite dispensing system for photon-pushed aerographite pellets remains as future work, as does the pellet-pusher. The pellet pusher is interesting in its own right as a means of producing hypervelocity laboratory particles or plasmas.

The requirement to maintain an overall efficiency of power extraction and pellet acceleration of greater than $\approx$80\% in order to obtain a significant velocity gain from wind--pellet shear sailing is of critical importance to the feasibility of the wind--pellet shear sailing concept. There are existing technologies, however, that may provide cause for optimism.  The Compact Linear Collider (CLIC), a concept being extensively explored for the next generation of particle accelerator, uses a low-energy, high-current particle accelerator as a \emph{drive beam} \cite{Stapnes2019}. More than 100~MW of power is extracted from this beam and transferred via a copper waveguide to a second accelerator---with a voltage gradient exceeding 100~MV/m---to create a greater energy but lower current beam. The overall efficiency of the power transfer process from the drive beam to the main beam can be as great as 97\% or greater \cite{Braun2008}. As it relates to wind--pellet shear sailing, we can see the wind of ISM playing an analogous role of the drive beam and the accelerator of pellets corresponding to the main beam of the CLIC. By using superconducting cable and accelerator tubes to process the extracted power, high efficiencies coupled with the extreme power levels required for wind--pellet shear sailing might be possible.

A further step along this line of development would be to use energetic pellets (such as fusion fuel pellets) rather than inert matter. The fusion-pellet runway idea of Kare \cite{Greason2020a} was a step in that direction but is limited by the drag of overtaking the pellets. There seems no reason in principle why wind--pellet shear sailing, which gains energy by overtaking pellets rather than losing energy, could not be increased in performance by pellets offering fusion energy content in addition to their kinetic energy. We chose not to include those effects in this paper because there are challenging implementation issues in recovering that energy at high overtaking speed and because we wished to focus on the underlying principle of gaining energy from shear sailing, but it is an interesting area for further inquiry.

\section{Conclusions}

In this study, the propulsive potential of using the velocity difference between the plasma medium in space (i.e., solar wind or ISM) and a stream of pellets, launched ``for free'' using either the photons or solar wind streaming outward from the sun, has been demonstrated. The principal novelty of the wind--pellet shear sailing concept is its ability to utilize pellets traveling slower than the velocity of the spacecraft, in contrast to earlier pellet stream and particle stream propulsion concepts proposed for interstellar flight. By using power extracted from the medium flowing past the spacecraft to launch the pellets backwards, acting as reaction mass as they are overtaken by the spacecraft, a fixed observer sees the kinetic energy of the pellets effectively being concentrated into the increased velocity of the spacecraft. A number of technical implementations of the power extraction device (i.e., windmill) and pellet pusher (i.e., accelerator) appear feasible, and the appearance of a technical roadblock for any particular technology necessary for power extraction from the ISM or for acceleration of the pellets should not invalidate the merit in further exploration of this concept. The efficiency of the power extraction from the apparent wind blowing past the spacecraft and its transfer to accelerating the pellets is of paramount concern in evaluating possible implementations. If realizable, the wind--pellet shear sailing approach appears particularly well suited to bridging the gap between velocities accessible via advanced photon sail or magnetic sail concepts, which reach a maximum of 0.02$c$, and the \emph{q}-drive concept, which becomes increasingly attractive if a spacecraft can achieve velocities of 0.05$c$ before engaging the \emph{q}-drive.

\section*{Acknowledgments}

The authors wish to thank the Interstellar Research Group, which sponsored the Tennessee Valley Interstellar Workshop that brought them together to begin these discussions. Thanks to Malik K. (MatterBeam) for asking a useful question at just the right time. The reviewers are thanked for examining and providing valuable feedback regarding the wind--pellet shear propulsion concept. Max Greason is thanked for providing assistance with the illustrations. The picture of a frozen water bubble is courtesy Ulrike Leone via Pixabay and used in accordance with their license. D.Y. and M.L. were supported by the Natural Sciences and Engineering Research Council of Canada (NSERC) and the McGill Summer Undergraduate Research in Engineering program.

\section*{Appendix A: Nondimensionalization of equation of motion}\label{sec:AppendixA}
\setcounter{figure}{0}
\label{sec:NondimensionalAppendix}
Since the equations of motion that govern the spacecraft acceleration are nonlinear ordinary differential equations that are not amenable to analytic solution (i.e., Eqs. (\ref{eq:FiniteISMode}) and (\ref{eq:SCframeEfficienciesODE})), it is convenient to nondimensionalize the equations such that they can be solved numerically in terms of fewer parameters.

The spacecraft velocity is nondimensionalized by the pellet velocity
\begin{eqnarray}
\phi = \frac{v_\mathrm{sc}}{v_\mathrm{p}}
\end{eqnarray}
where $v_\mathrm{p}$ is taken as a constant. The mass of pellets is nondimensionalized by the mass of the spacecraft
\begin{equation}
\xi = { \frac{m_\mathrm{p}}{m_\mathrm{sc}} }.
\end{equation}
With these nondimensionalizations, the governing differential equation can be written as
\begin{equation}
\alpha \left ( \frac{\mathrm{d} \phi}{\mathrm{d} \xi } \right )^2 + 2 \left ( \phi - \alpha \right ) \frac{\mathrm{d} \phi}{\mathrm{d} \xi } - (1 - \alpha) = 0 .
\end{equation}
If the efficiency of power extraction from the wind and pellet acceleration is included, the equation is
\begin{equation}
\alpha \left ( \frac{\mathrm{d} \phi}{\mathrm{d} \xi } \right )^2 + 2 \left ( \phi - \alpha \right ) \frac{\mathrm{d} \phi}{\mathrm{d} \xi } - 2 \phi \left ( 1 - \frac{1}{\eta} \right ) - \left ( \frac{1}{\eta} - \alpha \right )= 0 .
\end{equation}
These equations are solved numerically via Mathematica \textbf{NDSolve} subject to the initial conditions $\phi = 1$ at $\xi = 0$ and plotted for various values of $\alpha$ in Fig.~\ref{fig:VelocityCurves}(a) and values of $\eta$ in Fig.~\ref{fig:VelocityCurves}(b).

\section*{Appendix B: Analytical solution for the case of large ISM to pellet mass ratio}\label{sec:AappendixB}
\setcounter{figure}{0}

Solving Eq.~\ref{eq:FiniteISMode} for $\frac{\mathrm{d}v_\mathrm{sc}}{\mathrm{d}x}$ using the quadratic formula
\begin{equation}
 \frac{ \mathrm{d} v_\mathrm{sc} }{ \mathrm{d} m_\mathrm{p} } = 
 \frac{-\left ( v_\mathrm{sc} - \alpha \, v_\mathrm{p} \right ) \pm v_\mathrm{sc} \sqrt{1 - \left ( 2 - \frac{v_\mathrm{p}}{v_\mathrm{sc}} \right )\frac{v_\mathrm{p}}{v_\mathrm{sc}} \, \alpha } }{ \alpha \, m_\mathrm{sc} } .
\end{equation}
Only the plus sign of the $\pm$ corresponds to physical solutions. To avoid imaginary solutions, we would require
\begin{equation}
\left ( 2 - \frac{v_\mathrm{p}}{v_\mathrm{sc}} \right ) \frac{v_\mathrm{p}}{v_\mathrm{sc}} \, \alpha < 1
\end{equation}
which for the starting velocity $v_\mathrm{sc_i}=v_\mathrm{p}$ corresponds to the requirement that $\alpha < 1$, such that the ISM mass must exceed the pellet mass for physical solutions.

If $\alpha$ is small, then a Taylor series expansion can be used to approximate the square root term as $\sqrt{1-\epsilon} \approx 1 - \frac{\epsilon}{2} - \frac{\epsilon^2}{8} - \ldots$ and the equation of motion becomes
\begin{equation}
m_\mathrm{sc} \, \frac{ \mathrm{d} v_\mathrm{sc} }{ \mathrm{d} m_\mathrm{p} } \approx \frac{1}{2} \left [ 1 - \frac{\alpha}{4} \left ( 2 - \frac{v_\mathrm{p}}{v_\mathrm{sc}} \right)^2 \right ] \frac{v_\mathrm{p}^2}{v_\mathrm{sc}}.
\end{equation}
In the limit where the spacecraft velocity greatly exceeds the pellet velocity ($v_\mathrm{sc} \gg v_\mathrm{p}$),
\begin{equation}
m_\mathrm{sc} \, v_\mathrm{sc} \, \frac{ \mathrm{d} v_\mathrm{sc} }{ \mathrm{d} m_\mathrm{p} } \approx \frac{1}{2} \left ( 1 - \alpha \right ) \, v_\mathrm{p}^2
\end{equation}
permitting a separation of variables and solution
\begin{equation}\label{eq:SmallAlphaLimit}
\frac{1}{2} \, m_\mathrm{sc} \left ( v_\mathrm{sc_\mathrm{f}}^2 - v_\mathrm{sc_\mathrm{i}}^2 \right ) \approx \frac{1}{2} \left ( 1 - \alpha \right ) m_\mathrm{p_{tot}} \, v_\mathrm{p}^2.
\end{equation}
For $v_\mathrm{sc_\mathrm{i}} = v_\mathrm{p}$, this simplifies to
\begin{equation}\label{eq:FinalVelocitySmallAlphaAppendix}
v_\mathrm{sc_\mathrm{f}} \approx v_\mathrm{p} \sqrt{\left ( 1 - \alpha \right ) \frac{m_\mathrm{p_{tot}}}{m_\mathrm{sc}} + 1}.
\end{equation}
This approximate, analytic result (dashed curves) is compared to the numerical solution of the exact equation (solid curves) Eq.~(\ref{eq:FiniteISMode}) result in Fig.~\ref{fig:VelocityCurves}(a).


\bibliographystyle{elsarticle-num-names} 
\bibliography{WindPelletShear}





\end{document}